\documentclass[twocolumn,showpacs,preprintnumbers,amsmath,amssymb]{revtex4}

\usepackage{graphicx}
\usepackage{dcolumn}
\usepackage{bm}

\usepackage{color}
\usepackage[overlay,absolute]{textpos}
\definecolor{mygray}{gray}{0.4}
\definecolor{mylink}{rgb}{0.2, 0.2, 0.5}
\usepackage[colorlinks=true,urlcolor=mylink,dvipdfm,bookmarksnumbered=true,%
pdfborder={0 0 0},pdfpagemode=UseNone,pdfstartview=FitH]{hyperref}

\begin{document}

\begin{textblock*}{\textwidth}[0,0](19mm,11.5mm)
\footnotesize\noindent
\begin{minipage}{\textwidth}
\center
\textcolor{mygray}{Journal link:}
\href{http://dx.doi.org/10.1103/PhysRevA.75.023806}{http://dx.doi.org/10.1103/PhysRevA.75.023806}
\end{minipage}
\end{textblock*}

\begin{textblock*}{0.6\textwidth}[0,0](19mm,261mm)
\footnotesize\noindent
\begin{minipage}{\textwidth}
\textcolor{mygray}{Journal ref:} \href{http://dx.doi.org/10.1103/PhysRevA.75.023806}{A. E. B. Nielsen and K. M{\o}lmer, Phys.\ Rev.\ A \textbf{75}, 023806 (2007)}.\\
\textcolor{mygray}{Copyright (2007) by the American Physical Society.}
\end{minipage}
\end{textblock*}

\title{Single photon state generation from a continuous-wave non-degenerate
optical parametric oscillator}

\author{Anne E. B. Nielsen and Klaus M\o lmer}
\affiliation{Lundbeck Foundation Theoretical Center for Quantum
System Research, Department of Physics and Astronomy, University of
Aarhus, DK-8000 \AA rhus C, Denmark}

\begin{abstract}
We present a theoretical treatment of conditional preparation of
one-photon states from a continuous-wave non-degenerate optical
parametric oscillator. We obtain an analytical expression for the
output state Wigner function, and we maximize the one-photon state
fidelity by varying the temporal mode function of the output state.
We show that a higher production rate of high fidelity Fock states
is obtained if we condition the outcome on dark intervals around
trigger photo detection events.
\end{abstract}


\maketitle

\section{Introduction}

Light has come to play an important role in quantum information
sciences, and this puts focus on the ability to prepare and
manipulate state vectors of light, i.e., to restrict the field to a
few modes or a single field mode, and to interact exclusively with
the selected modes. In optical and microwave cavities, many
experiments have already been done on pure quantum state
manipulation in isolated field modes interacting with single atoms,
but there is also an obvious interest to produce traveling light
fields with controllable pure state properties.

It is not easy to manipulate quantum states of light, and one very
successful strategy has thus been to produce a certain "easy" state,
and to use a quantum measurement to project this state on the
desired quantum state. Following the theoretical proposals by Dakna
et al \cite{dakna}, see also \cite{sasaki,kim}, conditioned on
appropriate detection events, single- and two-mode squeezed states
of light can be transformed into Fock states and Schr\"odinger
cat-like states, as demonstrated with good fidelity with single mode
pulses of light in \cite{lvovsky,zavatta,grangier2,grangiercat}.
I.e., a large number of pulses are generated, and measurements,
carried out on each of them with some success probability, ascertain
the preparation of the desired state.

\begin{figure}
\begin{center}
\includegraphics*[viewport=5 5 155 97,width=0.90\columnwidth]{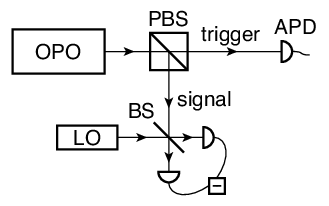}
\caption{Experimental setup for one-photon state production using a
type II OPO. A polarizing beam splitter (PBS) separates the source
output into two twin beams with opposite polarizations. One beam is
used as a trigger, and when the trigger detector (APD) registers a
photon, the generated state is investigated by homodyne detection
with a local oscillator (LO).} \label{setup}
\end{center}
\end{figure}

A scheme to produce single photon states and other single mode
quantum states from a continuous-wave field was demonstrated in
experiments \cite{jonas,wakui}, where a photon counter registered
the intensity of a small fraction of a beam of squeezed light,
causing the remaining beam to have a high single photon amplitude in
a well localized mode. A theoretical analysis of that experiment was
given in \cite{klaus}. In the present paper we generalize the
approach of \cite{klaus} to protocols involving twin beams of light
generated from a non-degenerate optical parametric oscillator (OPO).
The non-degenerate OPO converts photons from a pump beam into pairs
of photons in a pair of output modes, and a conditional detection of
a single photon in one beam (which we shall denote the trigger beam)
results in the presence of a single photon in the other beam (the
signal beam). A possible experimental setup is sketched in figure
\ref{setup}. In contrast to the pulsed case where the field mode
occupied is governed by the pulse shape, in the continuous-wave
case, the precise temporal modes occupied by the one-photon states
in the signal beam have to be specified. The largest one-photon
state fidelities are obtained by choosing the modes optimally, and
we present results of a variational procedure and of a numerical
optimization.

In Sec.\ II, we introduce the quantum correlation functions of the
twin beams of interest, and we show how the (Schr\"odinger picture)
quantum state populating an arbitrary field mode can be readily
obtained from the (Heisenberg picture) field correlation functions.
In Sec.\ III, we present analytical results for the phase space
Wigner function for the signal state conditioned on a single photo
detection event and for the fidelity and rate with which the
one-photon states are produced. In Sec.\ IV, we utilize a
variational method to optimize the signal mode function. Finally, in
Sec.\ V, we analyze the signal state conditioned on a photo
detection event surrounded by an interval with no photo detection
events. Sec.\ VI concludes the paper.

\section{Two-time quantum correlation functions and phase space
Wigner functions for Gaussian states}

The OPO Hamiltonian is quadratic in field annihilation and creation
operators, and the solution of the time evolution is most easily
accomplished by a linear (Bogoliubov) transformation of each pair of
incident field operators at all frequency components. The state
therefore retains the Gaussian state character of the vacuum state,
i.e., the joint probability distributions of all the quadrature
variables is Gaussian. The state separates in pairs of frequency
modes for which the quantum state can be written down explicitly.
Since we shall deal with photo detection experiments, we need a
representation of the field in time domain, and although it is still
a Gaussian state, it is now much more complicated, as the field
operators at different times are non-trivially correlated.

In so-called Type I and Type II OPO's, photons are created pairwise
in two different frequency components and in two different
polarization components, respectively. It is also possible to create
photons pairwise in beams propagating in different directions.
Alternatively one can combine the single beam squeezed light outputs
from two degenerate OPO's on a beam splitter, which also produce
twin correlated beams, if the OPO outputs are squeezed $\pi/2$ out
of phase. All of these approaches, if restricted to correlations in
only one degree of freedom (polarization, frequency, direction),
lead to quantum correlations described by the same formal
correlation functions for the annihilation and creation operators
for the two components, denoted in the following by $+$ and $-$,
$\hat{a}_+(t)$, $\hat{a}_+^\dag(t)$, $\hat{a}_-(t)$, and
$\hat{a}_-^\dag(t)$, and are in quite precise models of the OPO's
given by \cite{drummond}
\begin{equation}\label{korrelation}
\begin{split}
&\langle\hat{a}_\pm(t)\hat{a}_\mp(t')\rangle=\frac{\lambda^2-\mu^2}{4}
\left(\frac{\textrm{e}^{-\mu|t-t'|}}{2\mu}+
\frac{\textrm{e}^{-\lambda|t-t'|}}{2\lambda}\right)\\
&\langle\hat{a}_\pm^\dag(t)\hat{a}_\pm(t')\rangle=\frac{\lambda^2-\mu^2}{4}
\left(\frac{\textrm{e}^{-\mu|t-t'|}}{2\mu}-
\frac{\textrm{e}^{-\lambda|t-t'|}}{2\lambda}\right)\\
&\langle\hat{a}_\pm(t)\hat{a}_\pm(t')\rangle=
\langle\hat{a}_\pm^\dag(t)\hat{a}_\mp(t')\rangle=0
\end{split}
\end{equation}
where
\begin{equation}
\lambda=\frac{\gamma}{2}+\epsilon\quad\textrm{and}\quad\mu=
\frac{\gamma}{2}-\epsilon.
\end{equation}
The parameters in these expressions are the non-linear gain
coefficient $\epsilon$ of the OPO, and the decay rate $\gamma$ of
light in the OPO cavity due to leakage through the output mirror. It
is assumed that there is no loss through the other mirror. For fixed
$\gamma$ we note that $\epsilon/\gamma$ is closely related to the
mean twin beam intensity
$\langle\hat{a}_\pm^\dag(t)\hat{a}_\pm(t)\rangle=
2(\epsilon/\gamma)^2\gamma/(1-4(\epsilon/\gamma)^2))$.

The modes relevant to the experiment are the mode $f_1(t)$, in which
the trigger detection takes place, and the mode $f_2(t)$ occupied by
the produced state, while all other modes are unobserved. We assume
the trigger detection to take place on a timescale much shorter than
$\gamma^{-1}$. The precise shape of the detected temporal mode is
then irrelevant, and we assume
\begin{equation}\label{triggermodefunc}
f_1(t)=\left\{\begin{array}{cl}\frac{1}{\sqrt{\Delta
t_c}}&\textrm{if}\quad
t_c-\frac{\Delta t_c}{2} <t\leq t_c+\frac{\Delta t_c}{2}\\
0&\textrm{otherwise}\end{array}\right.,
\end{equation}
where $\Delta t_c$ is a short time interval and $t_c$ is the click
time for the trigger detection. As discussed in the introduction,
the mode function $f_2(t)$ can be chosen arbitrarily (under the
constraint $\int|f_2(t)|^2\mathrm{d}t=1$), and in particular it can
be optimized to achieve maximal fidelity. The optimal mode function
is centered around the trigger detector click time, and from
\eqref{korrelation} it is to be expected that the temporal extent is
of order $\gamma^{-1}$, which is the time that a photon can spend in
the cavity and hence be separated from its partner in the output
fields. We will return to this point in sections III and IV, but by
now we leave $f_2(t)$ unspecified. The problem thus reduces to the
one of characterizing the correlations of the single mode operators
\setlength\arraycolsep{2pt}
\begin{eqnarray}\label{modeope}
\hat{a}_1&=&\int f_1(t')\Big(\sqrt{\eta_t}\hat{a}_+(t')
+\sqrt{1-\eta_t}\hat{a}_{+,vac}(t')\Big)\mathrm{d}t',\\
\label{modeope2} \hat{a}_2&=&\int
f_2(t')\Big(\sqrt{\eta_s}\hat{a}_-(t')
+\sqrt{1-\eta_s}\hat{a}_{-,vac}(t')\Big)\mathrm{d}t',
\end{eqnarray}
where $\eta_t$ is the trigger detector efficiency, $\eta_s$ is the
signal detector efficiency, and $\hat{a}_{\pm,vac}$ are field
operators acting on vacuum, included to ensure appropriate
commutator relations of $\hat{a}_i$ and $\hat{a}^\dag_i$.

Since the state is Gaussian, we immediately know the multi-component
Wigner function of the trigger and signal mode before conditioning
on the trigger detector click event. Defining the column vector
\begin{equation}
y=(x_1,p_1,x_2,p_2)^T,
\end{equation}
of quadrature variables
$\left(\hat{a}_i=(\hat{x}_i+i\hat{p}_i)/\sqrt{2}\right)$, this
Wigner function is
\begin{equation}\label{wv}
W_V(y)=\frac{1}{\pi^2\sqrt{\det(V)}}\exp\left(-y^TV^{-1}y\right).
\end{equation}
$W_V(y)$ is parameterized by the covariance matrix $V$ with elements
$V_{ij}=\langle \hat{y}_i\hat{y}_j\rangle+\langle
\hat{y}_j\hat{y}_i\rangle$, which are computed explicitly by use of
equations \eqref{modeope} and \eqref{modeope2}, the time dependent
mode functions $f_i(t)$, and the two-time correlation functions
\eqref{korrelation}. Since it follows from \eqref{korrelation} that
$\langle\hat{a}_1\hat{a}_1\rangle=\langle\hat{a}_2\hat{a}_2\rangle=
\langle\hat{a}^\dag_1\hat{a}_2\rangle=0$, we have
\begin{eqnarray}\label{V11}
V_{11}&=&V_{22}=1+2\langle\hat{a}_1^\dag\hat{a}_1\rangle,\\
V_{33}&=&V_{44}=1+2\langle\hat{a}_2^\dag\hat{a}_2\rangle,\\
V_{12}&=&V_{21}=V_{34}=V_{43}=0,\\
V_{13}&=&V_{31}=-V_{24}=-V_{42}=2\mathrm{Re}(\langle\hat{a}_1\hat{a}_2\rangle),\\
\label{V14}
V_{14}&=&V_{41}=V_{23}=V_{32}=2\mathrm{Im}(\langle\hat{a}_1\hat{a}_2\rangle).
\end{eqnarray}
We note that we use the same symbols for quadrature operators and
for the corresponding real variable arguments in the Wigner
function.

\section{Analysis of the state conditioned on the detection of a trigger photon}

One can apply different trigger detector models taking into account
the detailed physical functioning of the detector. In \cite{klaus},
the theory is outlined for three such models: a perfect photon
counter, an "on/off" detector that discriminates vacuum from
non-vacuum states, and a "click" detector based on the absorption of
a photon by the photoelectric effect. For general Gaussian states
there are significant differences between the outcomes of these
detector models, but as we are here interested in a trigger mode of
infinitesimal duration and hence with vanishing populations of
higher photon number states, they give identical results for the
signal field state after the final state of the trigger mode is
traced out. We henceforth apply the normal photo detector theory,
where a click detection is accompanied by the application of the
trigger field annihilation operator on the quantum state of the
system, i.e., by application of the annihilation operator from the
left and its adjoint creation operator from the right on the density
matrix. Afterwards, the otherwise unobserved trigger mode is traced
out.

\begin{figure}
\includegraphics*[viewport=9 10 400 299,width=0.95\columnwidth]{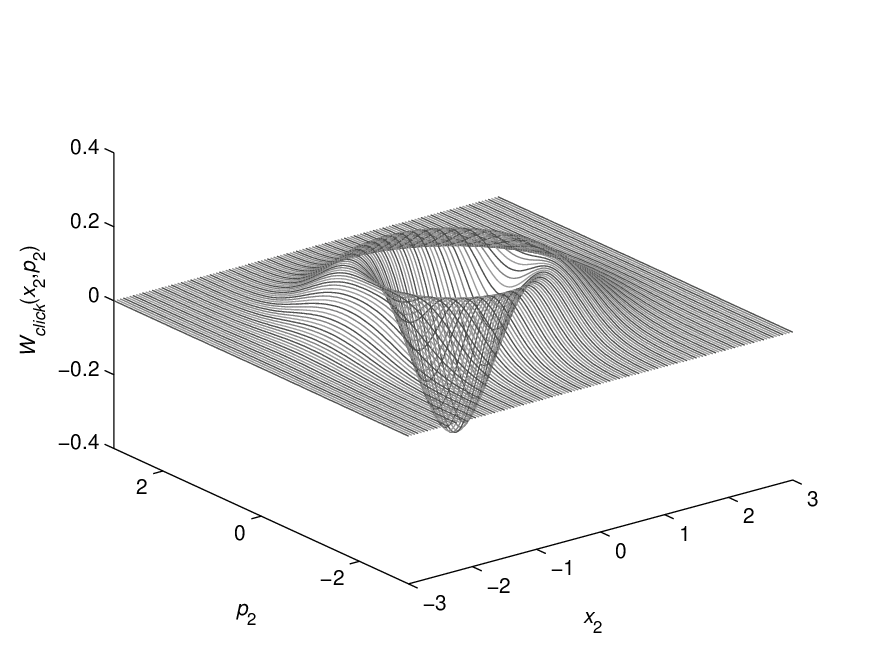}
\caption{Wigner function conditioned on a single trigger detector
click event for $\epsilon/\gamma=0.02$, $\eta_s=1$, and
$f_2(t)=\sqrt{\gamma/2}\exp(-\gamma|t-t_c|/2)$. The one-photon state
fidelity is 0.9921, and the negativity at the origin is
$W_{click}(0,0)=-0.3133$.}\label{wignerone}
\end{figure}

The application of the annihilation and creation operations is
mapped to differential operators on the Wigner function
\cite{gardiner}, and the partial trace is performed by an
integration over the corresponding phase space variables. This
results in the following Wigner function for the conditioned signal
state
\begin{multline}\label{Wclick}
W_{click}(x_{2},p_{2})=N_{click}\int \mathrm{d}x_1\mathrm{d}p_1
\frac{1}{2}
\bigg(1+x_1^2+p_1^2\\+\frac{1}{4}\left(\frac{\partial^2}{\partial
x_1^2}+\frac{\partial^2}{\partial
p_1^2}\right)+x_1\frac{\partial}{\partial
x_1}+p_1\frac{\partial}{\partial p_1}\bigg) W_V(y),
\end{multline}
where $N_{click}$ is a normalization constant. By partial
integration \eqref{Wclick} reduces to
\begin{eqnarray}
W_{click}(x_{2},p_{2})&=&N_{click}\int \mathrm{d}x_1\mathrm{d}p_1
\frac{1}{2}
\left(x_1^2+p_1^2-1\right)W_V(y)\nonumber\\
&=&\left(A_1+A_2(x_2^2+p_2^2)\right)\mathrm{e}^{-A_3(x_2^2+p_2^2)},
\end{eqnarray}
where the coefficients $A_1$, $A_2$, and $A_3$ are given in terms of
the covariance matrix elements (\ref{V11}-\ref{V14})
\begin{eqnarray*}
A_1&=&\frac{V_{11}V_{33}-V_{13}^2-V_{14}^2-V_{33}}{\pi(V_{11}-1)V_{33}^2},\\
A_2&=&\frac{V_{13}^2+V_{14}^2}{\pi(V_{11}-1)V_{33}^3},\\
A_3&=&(V_{33})^{-1}.
\end{eqnarray*}
$W_{click}$ depends on the signal mode function $f_2(t)$, the signal
detector efficiency $\eta_s$, and the non-linear gain coefficient
$\epsilon$ of the OPO through $V$, but $W_{click}$ does not depend
on the trigger detector efficiency $\eta_t$. Since we condition on a
click event, the only effect of an inefficient trigger detector is
to reduce the production rate. Figure \ref{wignerone} shows
$W_{click}$ for $\epsilon/\gamma=0.02$, $\eta_s=1$, and
$f_2(t)=\sqrt{\gamma/2}\exp(-\gamma|t-t_c|/2)$ (in the next section
we prove that this is the optimal mode function for
$\epsilon/\gamma\rightarrow0$). The fidelity, by which the
conditioned Wigner function $W_{click}$ resembles the Wigner
function for a one-photon state
\begin{equation}
W_{n=1}(x,p)=\pi^{-1}\left(-1+2(x^2+p^2)\right)\mathrm{e}^{-(x^2+p^2)},
\end{equation}
is
\begin{multline}\label{fidone}
F_1(f_2(t))\equiv2\pi\iint
W_{click}(x_2,p_2)W_{n=1}(x_2,p_2)\mathrm{d}x_2\mathrm{d}p_2\\
=\frac{2(V_{11}-1)(V_{33}^2-1)
+2(3-V_{33})(V_{13}^2+V_{14}^2)}{(V_{11}-1)(1+V_{33})^3},
\end{multline}
and for the Wigner function in figure \ref{wignerone} we find
$F_1(f_2(t))=0.9921$.

For completeness we present in appendix A analytical results for the
Wigner function and the one-photon state fidelity for the degenerate
OPO setup analyzed in \cite{klaus}.

\begin{figure}
\includegraphics*[viewport=9 7 400 299,width=0.95\columnwidth]{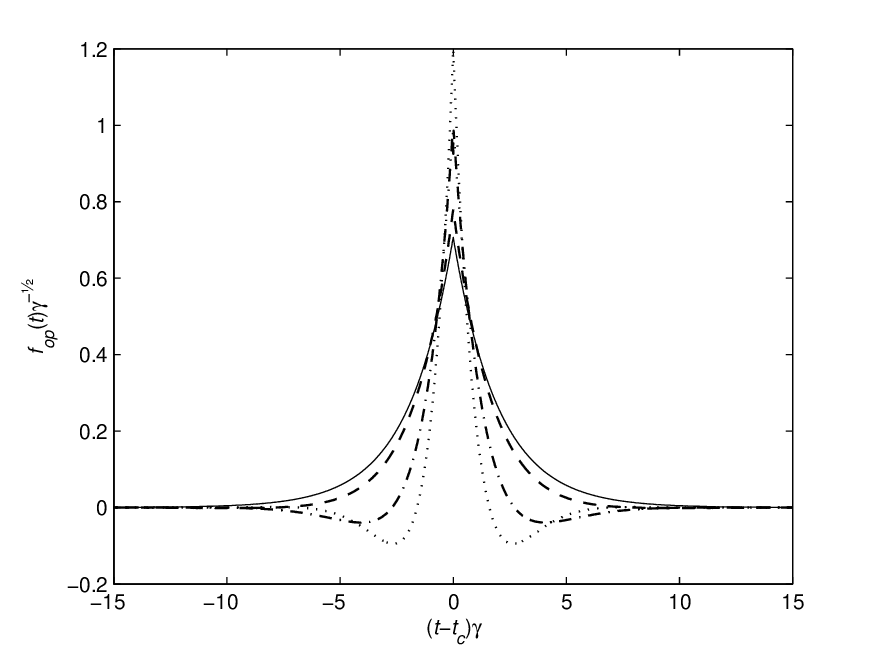}
\caption{Optimal mode function $f_{op}(t)$ for $\epsilon/\gamma=0$
(solid line), $\epsilon/\gamma=0.1$ (dashed line),
$\epsilon/\gamma=0.2$ (dot-dashed line), and $\epsilon/\gamma=0.3$
(dotted line). The signal detector efficiency is
$\eta_s=1$.}\label{modeone}
\end{figure}

The optimal signal mode function (denoted $f_{op}(t)$) obtained from
numerical optimization of the one-photon state fidelity
\eqref{fidone} is shown in figure \ref{modeone} for $\eta_s=1$ and
different values of $\epsilon/\gamma$. (The optimization is over all
real functions, which will be justified in the next section.) The
tendency is that the width of the mode function decreases, when the
beam intensity is increased, and a dip appears on each side of the
peak. However, the difference between the optimal mode function for
$\epsilon/\gamma=0.1$ and for $\epsilon/\gamma=0$ is quite small.

\begin{figure}
\begin{center}
\includegraphics*[viewport=18 9 388 299,width=0.95\columnwidth]{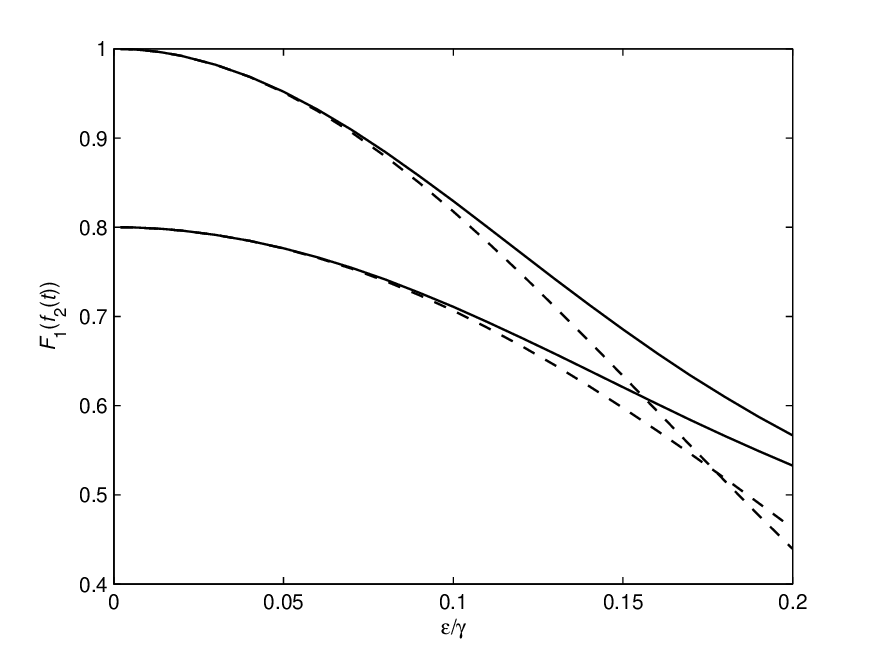}
\caption{Fidelity calculated from the numerically optimized mode
function (solid lines) and from the $\epsilon=0$ mode function in
equation \eqref{enoutmodefunc} (dashed lines). The curves
approaching 1 for small $\epsilon$ are for perfect detection
$\eta_s=1$, and $\eta_s=0.8$ for the curves approaching 0.8 for
small $\epsilon$} \label{entwinepsfid}
\end{center}
\end{figure}

The optimized fidelity is shown in figure \ref{entwinepsfid} for
$\eta_s=1$ and $\eta_s=0.8$ (solid lines). The fidelity decreases
with increasing $\epsilon$ because the two-photon state contribution
to the output state increases when the intensity increases. The
effect of reducing $\eta_s$ from 1 to 0.8 is roughly to reduce the
fidelity by 20 \% in the small $\epsilon/\gamma$ region. For larger
$\epsilon/\gamma$ the reduction is smaller because the lower
efficiency reduces the intensity seen by the detector. The dashed
curves in the figure show the fidelity calculated for the mode
function $f_2(t)=\sqrt{\gamma/2}\exp(-\gamma|t-t_c|/2)$, and it is
apparent that this analytical approximation is close to optimal for
$\epsilon/\gamma\lesssim 0.1$. This region is also the most
interesting region since the fidelity is low for larger
$\epsilon/\gamma$.

\begin{figure}
\begin{center}
\includegraphics*[viewport=18 9 388 299,width=0.95\columnwidth]{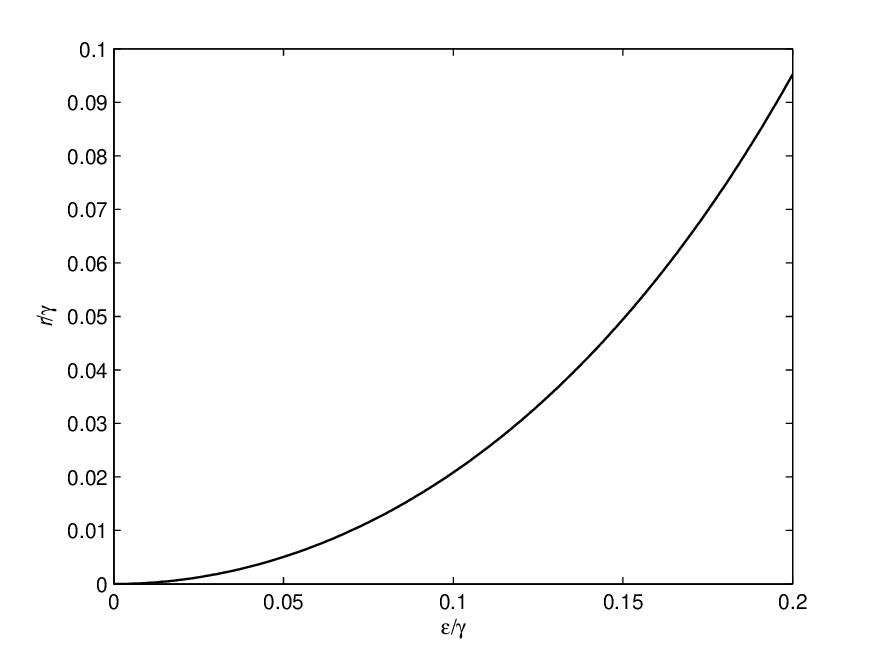}
\caption{Production rate $r$ versus $\epsilon/\gamma$ for
$\eta_t=1$.} \label{rate}
\end{center}
\end{figure}

The production rate $r$ (i.e. the trigger detector click rate) is
$r=P_{click}/\Delta t_c$, where $P_{click}$ is the probability to
observe a click in the trigger mode function. We assume that
$\langle\hat{a}_1^\dag\hat{a}_1\rangle<<1$, which is valid if
$\Delta t_c$ is small compared to the mean temporal distance between
the photons in the trigger beam. The probability to observe a click
in the trigger mode is then the expectation value of the number of
photons in that mode, and hence
\begin{equation}
r=\frac{\langle\hat{a}_1^\dag\hat{a}_1\rangle}{\Delta t_c}
=\frac{2(\epsilon/\gamma)^2} {1-4(\epsilon/\gamma)^2}\gamma\eta_t.
\end{equation}
Figure \ref{rate} shows the production rate as a function of
$\epsilon/\gamma$ for $\eta_t=1$. The rate decreases when
$\epsilon/\gamma$ decreases and is zero when $\epsilon/\gamma=0$.
However, to achieve a high fidelity, $\epsilon/\gamma$ needs to be
small according to figure \ref{entwinepsfid}. Assuming perfect
signal transmission, one-photon states with a fidelity of
$F_1(f_{op}(t))=0.95$ can be produced at rates of tens of kHz, if we
assume $\epsilon/\gamma=0.05$ and $\gamma$ of order
$5\cdot10^7\textrm{ s}^{-1}$. These numbers are in quantitative
accord with the experimental parameters and count rates for
degenerate OPO's \cite{jonas,wakui}. In Sec.\ V we analyze the
possibility to obtain better production rates and fidelities by
conditioning on clicks surrounded by intervals with no click
detection events, but first we consider the mode function
optimization in greater detail.

\section{Variational optimization of the signal mode function}

In this section we optimize the signal mode function by analytical
variation of the one-photon state fidelity \eqref{fidone}. We write
the signal mode function as
$f_2(t)=|f_2(t)|\mathrm{e}^{i\theta(t)}$, where $\theta(t)$ is real.
$|f_2(t)|$ and $\theta(t)$ may be varied independently, and since
the constraint $\int|f_2(t)|^2\mathrm{d}t=1$ does not involve
$\theta(t)$, the optimal choice of $\theta(t)$ fulfils that the
variation of $F_1$ vanishes, when $\theta(t)$ is varied, which leads
to
\begin{multline}\label{theta}
\int |f_2(t)||f_2(t')|\sin\left(\theta(t')-\theta(t)\right)
\Big(c_1(f_2)\langle\hat{a}_-^\dag(t)\hat{a}_-(t')\rangle\\+
c_2(f_2)\langle\hat{a}_+(t_c)\hat{a}_-(t)\rangle
\langle\hat{a}_+(t_c)\hat{a}_-(t')\rangle\Big)\mathrm{d}t'=0,
\end{multline}
where
\begin{eqnarray*}
c_1(f_2)&=&4\eta_s\bigg(\frac{1}{(1+V_{33})^2}
-\frac{V_{13}^2+V_{14}^2}{(V_{11}-1)(1+V_{33})^3}\\
&&\quad-\frac{2(V_{33}-1)}{(1+V_{33})^3}
-\frac{3(3-V_{33})(V_{13}^2+V_{14}^2)}{(V_{11}-1)(1+V_{33})^4}\bigg),\\
c_2(f_2)&=&\frac{8\eta_s\eta_t\Delta
t_c(3-V_{33})}{(V_{11}-1)(1+V_{33})^3}.
\end{eqnarray*}
Note that $c_1$ and $c_2$ depend on the signal mode function through
the covariance matrix elements, but they do not depend on time. Note
also that $V_{11}-1$ and $V_{13}^2+V_{14}^2$ are both proportional
to $\eta_t\Delta t_c$, while $V_{33}$ is independent of $\eta_t$ and
$\Delta t_c$.

For a general $|f_2(t)|$ the solution to \eqref{theta} is
$\theta(t)=constant+N\pi$, where $N$ is an integer that may be
different at different times. \eqref{theta} is also fulfilled in the
limit where $\theta(t)$ varies infinitely fast with time, but this
leads to $V_{33}=1$ and $V_{13}^2+V_{14}^2=0$, and hence $W_{click}$
reduces to the Wigner function for vacuum and $F_1=0$. Thus the
phase is constant except for possible sign shifts. Adding a time
independent constant to $\theta(t)$ is equivalent to rotating the
signal state Wigner function in phase space, but since the desired
$|n=1\rangle$ signal state Wigner function depends on $x_2^2$ and
$p_2^2$ through $x_2^2+p_2^2$, it is rotationally symmetric, and
hence the fidelity is unchanged by this transformation.
Consequently, we can choose $f_{op}(t)=\pm|f_{op}(t)|$, and it is
thus sufficient to optimize over all real functions.

Restricting $f_2(t)$ to be real, the constraint is $\int
f_2(t)^2\mathrm{d}t=1$, and this is taken into account by
introducing a Lagrange multiplier $\xi$ and demanding the variation
of $F_1-\xi\int f_2(t)^2\mathrm{d}t$ to vanish when $f_2(t)$ is
varied. This leads to the following integral equation for the
optimal signal mode function
\begin{multline}\label{intone}
\xi f_{op}(t)=\\
c_1(f_{op})\frac{\lambda^2-\mu^2}{4}\int f_{op}(t')
\left(\frac{\textrm{e}^{-\mu|t-t'|}}{2\mu}-
\frac{\textrm{e}^{-\lambda|t-t'|}}{2\lambda}\right)\mathrm{d}t'\\
+c_2(f_{op})\frac{V_{13}}{\sqrt{\eta_s\eta_t\Delta
t_c}}\frac{\lambda^2-\mu^2}{8}\left(\frac{\textrm{e}^{-\mu|t-t_c|}}{2\mu}+
\frac{\textrm{e}^{-\lambda|t-t_c|}}{2\lambda}\right).
\end{multline}
For $\epsilon=0$ the first term in \eqref{intone} vanishes, and we
have
\begin{equation}\label{enoutmodefunc}
\lim_{\epsilon\rightarrow0}f_{op}(t)=
\sqrt{\frac{\gamma}{2}}\exp\left(-\frac{\gamma}{2}|t-t_c|\right).
\end{equation}
\eqref{enoutmodefunc} is the optimal zero intensity signal mode
function, and for $\eta_s=1$ the fidelity is
$\lim_{\epsilon\rightarrow0}F_1(f_{op}(t))=1$. For $\epsilon\neq0$,
\eqref{enoutmodefunc} can be used as the starting point for a
numerical iteration of \eqref{intone}. To increase the range of
$\epsilon/\gamma$ values, where the iteration procedure converges,
we use the mode function
$N_\alpha((1-\alpha)(f_2(t))_i+\alpha(f_2(t))_{i-1})$ in the
$(i+1)^{\mathrm{th}}$ iteration step, where $(f_2(t))_i$ is the mode
function obtained from the $i^{\mathrm{th}}$ step. $\alpha$ is a
number between $0$ and $1$, and $N_\alpha$ is a normalization
constant. Using this method we regain the mode functions in figure
\ref{modeone}.

\section{Continuous detections with (efficient) trigger detector}

In Sec.\ III we calculated the one-photon state fidelity for the
output state conditioned on a single click event. Conditioning
solely on one click event means that we average over all possible
outcomes of the trigger detector measurements outside the small time
window where the conditioning click occurs. This method is correct
if the trigger detector is turned off outside the conditioning click
time window, or if we want to calculate the average fidelity for a
large number of produced states where no selection of particular
click events (for instance those with no other clicks nearby) has
taken place. However, an improvement in fidelity for fixed
$\epsilon/\gamma$ can be obtained if we only accept clicks that are
surrounded by a certain time period $T$ with no clicks since this
restricts the two-photon and higher photon number state
contributions to the produced state. The fidelity increase depends
on the trigger detector efficiency. If $\eta_t=1$, it is certain
that no photons hit the trigger detector during the period $T$ if no
clicks are registered, and for large $T\gamma$ the fidelity
approaches unity irrespective of the value of $\epsilon/\gamma$
(provided $\Delta t_c\gamma$ is much smaller than the mean temporal
distance between photons in the trigger beam). In the opposite
limit, $\eta_t=0$, a vacuum detection (i.e. no click) gives us no
information about the produced state, and we regain the results from
Sec.\ III.

For fixed $\epsilon/\gamma$ the conditioning on a no click interval
results in a smaller production rate because some clicks are now
disregarded, but since the fidelity is increased, we can increase
$\epsilon/\gamma$ a little while still obtaining a larger fidelity
than in Sec.\ III, and if $T\gamma$ is not too large, this leads to
a larger production rate. In the region where the rate is limited by
the number of photons produced by the source it thus turns out that
an increase in production rate for fixed fidelity or an increase in
fidelity for fixed production rate can be obtained.

In the following we first calculate the one-photon state fidelity
for the signal state conditioned on one click detection and an
arbitrary number of no click detections, and after that we derive an
equation for the corresponding production rate. Finally we present
and discuss numerical results for the fidelity and the production
rate.

\subsection{Fidelity}

To determine the one-photon state fidelity for the produced state
conditioned on a click event surrounded by a time interval with no
click events, we replace the continuous time argument by a discrete
set of box shaped temporal trigger mode functions and assume the
trigger detections to take place in these modes. The $m=T/\Delta
t_c$ vacuum detection modes are labeled by the numbers
$3,4,\ldots,m+2$ and are included in the covariance matrix. The
unconditioned Wigner function is then
\begin{equation}
W_V(y)=\frac{1}{\pi^{m+2}\sqrt{\det(V)}}\exp\left(-y^TV^{-1}y\right),
\end{equation}
where $y=(x_1,p_1,x_2,p_2,\ldots,x_{m+2},p_{m+2})^T$. We first
determine the Wigner function for the state conditioned on all the
vacuum detections but not the click. A vacuum detection results in a
projection of the relevant mode on the vacuum state, which in terms
of Wigner functions corresponds to multiplication with the vacuum
state Wigner function $W_{n=0}(x,p)=\pi^{-1}\exp(-x^2-p^2)$ followed
by integration over the relevant quadrature variables and
renormalization. Thus the Wigner function $W_{vaccon}$ for the state
conditioned on the $m$ vacuum detections is
\begin{multline}
W_{vaccon}(x_1,p_1,x_2,p_2)=\\N_{vaccon}\bigg(\prod_{i=3}^{m+2}\int
\mathrm{d}x_i\mathrm{d}p_i W_{n=0}(x_i,p_i)\bigg)W_V(y),
\end{multline}
where $N_{vaccon}$ is a normalization constant. Since the vacuum
projection is a gaussian operation, $W_{vaccon}$ is also a gaussian
function. This means that we can proceed as in Sec.\ III if we just
replace the covariance matrix for the output mode and the click mode
with the covariance matrix $V_{vaccon}$ corresponding to the
gaussian Wigner function $W_{vaccon}$. If we write the
$(2m+4)\times(2m+4)$ covariance matrix $V_{2m+4}$ for the signal
mode, the click mode, and the vacuum modes as
\begin{equation}
V_{2m+4}=\left[\begin{array}{cc}V_4&C\\C^T&V_{2m}\end{array}\right],
\end{equation}
where $V_4$ is the $4\times4$ covariance matrix for the signal mode
and the click mode, and $V_{2m}$ is the covariance matrix for the
vacuum modes, it has been proven \cite{eisert} that
\begin{equation}\label{vvaccon}
V_{vaccon}=V_4-C(V_{2m}+I_{2m})^{-1}C^T,
\end{equation}
where $I_{2m}$ is the $(2m)\times(2m)$ identity matrix. Hence by use
of \eqref{vvaccon} we can immediately calculate the improved
fidelity from \eqref{fidone} (note that the covariance matrix
elements that are equal in Sec.\ III are changed by equal amounts
and are hence still equal).

\subsection{Production rate}

The production rate is
\begin{equation}\label{r}
r=\frac{P_{vac,click}}{\Delta t_c},
\end{equation}
where $P_{vac,click}$ is the probability to detect no click in the
$m$ vacuum modes and a click in the click mode. To determine
$P_{vac,click}$ we calculate the overlap of the unconditioned source
output state with the vacuum state in the $m$ vacuum modes and then
the expectation value of
$\hat{a}_1^\dag\hat{a}_1=\frac{1}{2}(\hat{x}_1^2+\hat{p}_1^2-1)$ in
the click mode
\begin{multline}\label{pvc}
P_{vac,click}=(2\pi)^m\int
\frac{1}{2}\left(x_1^2+p_1^2-1\right)\\
\left(\prod_{i=2}^{m+1}W_{n=0}(x_i,p_i)\right)W_V(y)\mathrm{d}y.
\end{multline}
Since we consider trigger modes only, the signal mode is not
included in the $(2m+2)\times(2m+2)$ covariance matrix in
\eqref{pvc}. Performing the integration we find
\begin{multline}\label{pvc2}
P_{vac,click}=\frac{2^{m-1}}{\sqrt{\det(I_{2m+2}+VJ)}}\\
\left(\frac{\det\left((V_x^{-1}+J_x)_{red}\right)}{\det(V_x^{-1}+J_x)}-1\right),
\end{multline}
where we have introduced the following notation: $I_{2m+2}$ is the
$(2m+2)\times(2m+2)$ identity matrix, $J$ is a $(2m+2)\times(2m+2)$
matrix with elements $J_{ii}=1$ for $i>2$ and $J_{ij}=0$ otherwise,
$V_x$ is the $(m+1)\times(m+1)$ matrix consisting of all the odd-odd
matrix elements of $V$, $J_x$ is an $(m+1)\times(m+1)$ matrix with
elements $(J_x)_{ii}=1$ for $i>1$ and $(J_x)_{ij}=0$ otherwise, and
$red$ means that the first row and the first column of the matrix
(corresponding to the click mode) have been removed.

\subsection{Numerical results}

In all the numerical calculations in this subsection $\eta_s=1$ and
$\Delta t_c\gamma=0.02$. For $\epsilon/\gamma=0.2$ (and $\eta_t=1$)
we find $\langle\hat{a}_1^\dag\hat{a}_1\rangle=2\cdot10^{-3}$, and
hence the assumption $\langle\hat{a}_1^\dag\hat{a}_1\rangle<<1$ is
valid for the region, where the fidelity is large. Even for
$\epsilon/\gamma=0.45$,
$\langle\hat{a}_1^\dag\hat{a}_1\rangle=4\cdot10^{-2}$ is still
somewhat smaller than $1$.

\begin{figure}
\begin{center}
\includegraphics*[viewport=7 7 388 296,width=0.95\columnwidth]{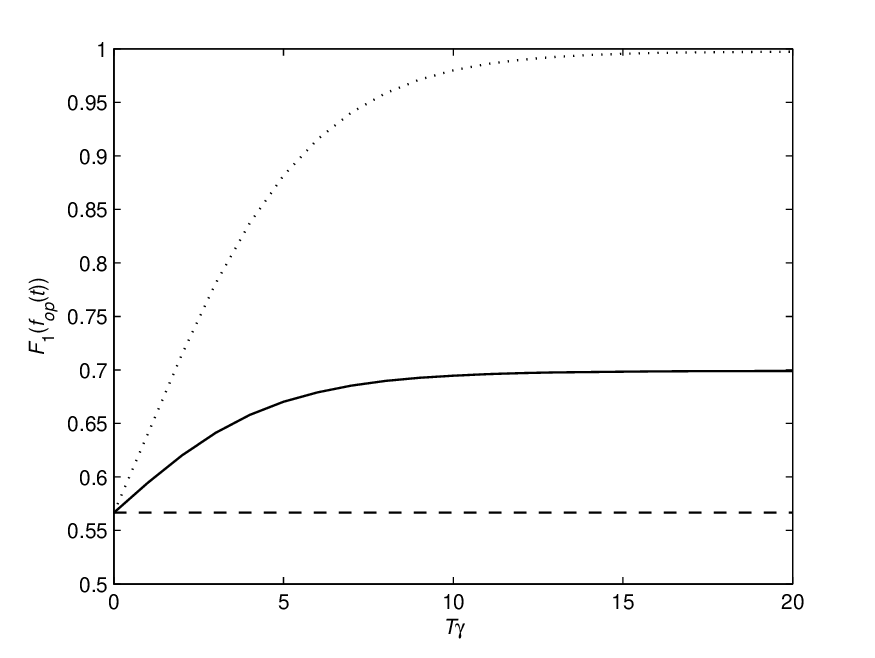}
\caption{Optimized one-photon fidelity as a function of vacuum
detection window $T$. The trigger detector efficiency is $\eta_t=1$
for the dotted line, $\eta_t=0.4$ for the solid line, and $\eta_t=0$
for the dashed line, while $\epsilon/\gamma=0.2$ for all curves.}
\label{femTfid}
\end{center}
\end{figure}

\begin{figure}
\begin{center}
\includegraphics*[viewport=7 7 388 296,width=0.95\columnwidth]{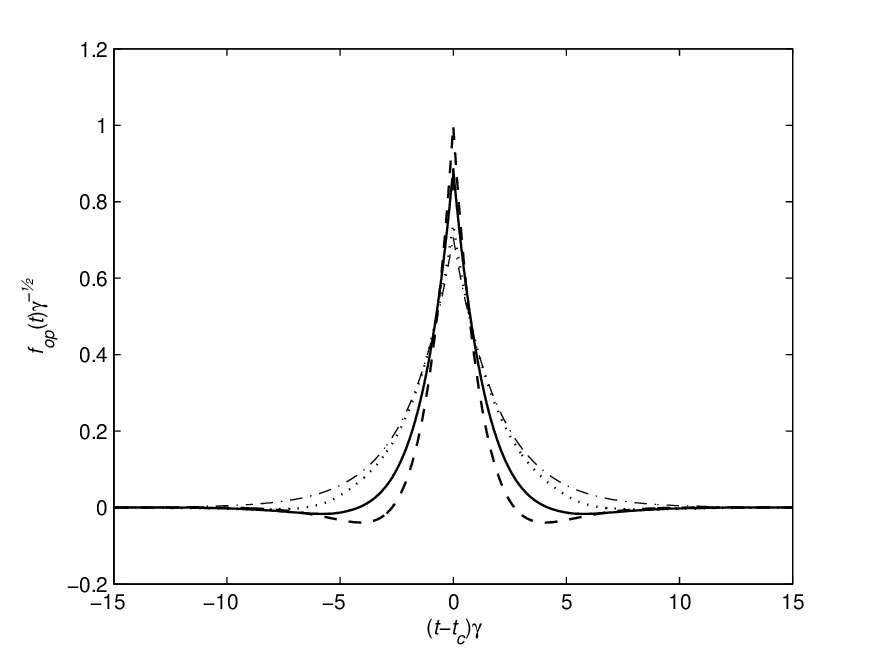}
\caption{Optimal mode function for $\eta_t=1$ (dotted line),
$\eta_t=0.4$ (solid line), and $\eta_t=0$ (dashed line).
$\epsilon/\gamma=0.2$ and $T\gamma=10$. The dot-dashed line is the
$\epsilon/\gamma=0$ optimal mode function given by
\eqref{enoutmodefunc}.} \label{femmode}
\end{center}
\end{figure}

Figure \ref{femTfid} shows the optimized one-photon state fidelity
as a function of the time interval $T$ in which we demand no click
events to occur. $T$ is chosen symmetrically around the trigger
detector click time, i.e. we condition on no clicks from
$t=t_c-\Delta t_c/2-T/2$ to $t=t_c-\Delta t_c/2$ and from
$t=t_c+\Delta t_c/2$ to $t=t_c+\Delta t_c/2+T/2$, since this gives
rise to the largest fidelity increase for a given $T\gamma$. The
rate and fidelity expressions are, however, also valid for
asymmetrical time intervals (and even for time intervals interrupted
with periods where the trigger detector is turned off). As expected
the fidelity increases when $T\gamma$ increases if $\eta_t\neq0$,
and for $\eta_t=1$ the fidelity approaches one for large $T\gamma$.
For a moderate trigger detector efficiency $\eta_t=0.4$, the
fidelity increase is smaller, but still significant. The fidelity
increase levels off when $T\gamma$ increases beyond approximately
$10$ irrespective of the value of $\eta_t$. The reason for this is
that the temporal extent of the signal mode function is
approximately $10\textrm{ }\gamma^{-1}$ as is apparent from figure
\ref{femmode}, which shows the optimal signal mode function for
$T\gamma=10$, $\epsilon/\gamma=0.2$, and different trigger detector
efficiencies. For $\eta_t=0$ the mode function is identical to the
$\epsilon/\gamma=0.2$ mode function in figure \ref{modeone}, but
when $\eta_t$ is increased to non-zero values, the mode function
approaches the optimal $\epsilon/\gamma=0$ mode function
\eqref{enoutmodefunc} (the dot-dashed line in the figure).

\begin{figure}
\begin{center}
\includegraphics*[viewport=7 7 388 296,width=0.95\columnwidth]{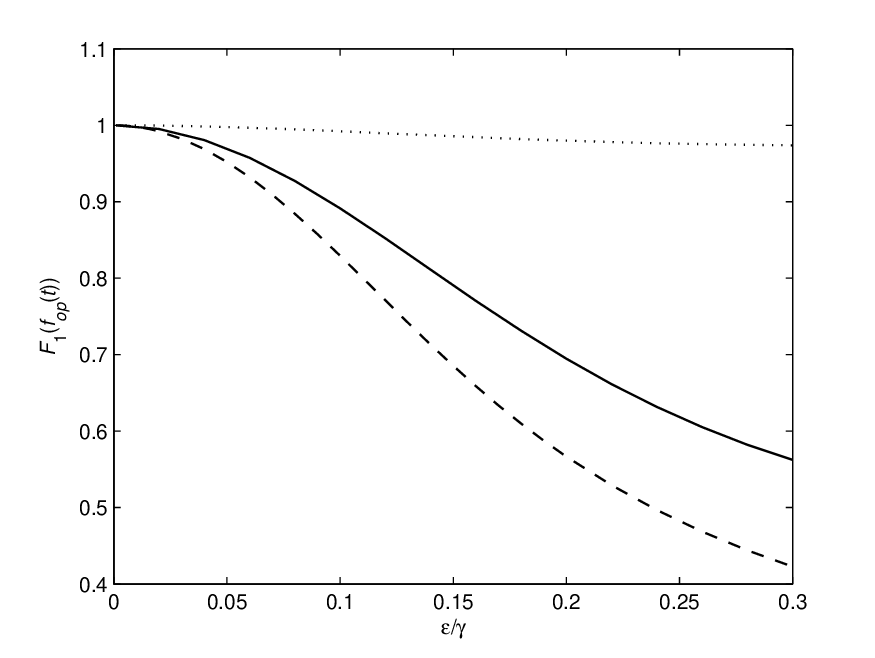}
\caption{Maximal one-photon fidelity versus $\epsilon$ for
$T\gamma=10$. The dotted line corresponds to $\eta_t=1$, while
$\eta_t=0.4$ for the solid line. The dashed line is for $\eta_t=0$,
and it is hence identical to the upper (solid) curve in figure
\ref{entwinepsfid}.} \label{femepsfidid}
\end{center}
\end{figure}

The optimized one-photon fidelity as a function of $\epsilon$ for
$T\gamma=10$ and different values of trigger detector efficiency is
plotted in figure \ref{femepsfidid}. The dashed curve ($\eta_t=0$)
is identical to the $T=0$ curve in Sec.\ III, so the difference
between the solid line and the dashed curve shows the fidelity
increase $\Delta F_1$ when $T\gamma$ increases from 0 to 10 for
$\eta_t=0.4$. $\Delta F_1$ increases with $\epsilon$ because the
mean temporal distance between the photons in the trigger and the
signal beam decreases when $\epsilon/\gamma$ increases, and hence it
is more likely to have close clicks. The dotted curve in the figure
is for perfect trigger detection $\eta_t=1$, and as expected it is
close to unity for all $\epsilon/\gamma$.

\begin{figure}
\begin{center}
\includegraphics*[viewport=7 7 388 296,width=0.95\columnwidth]{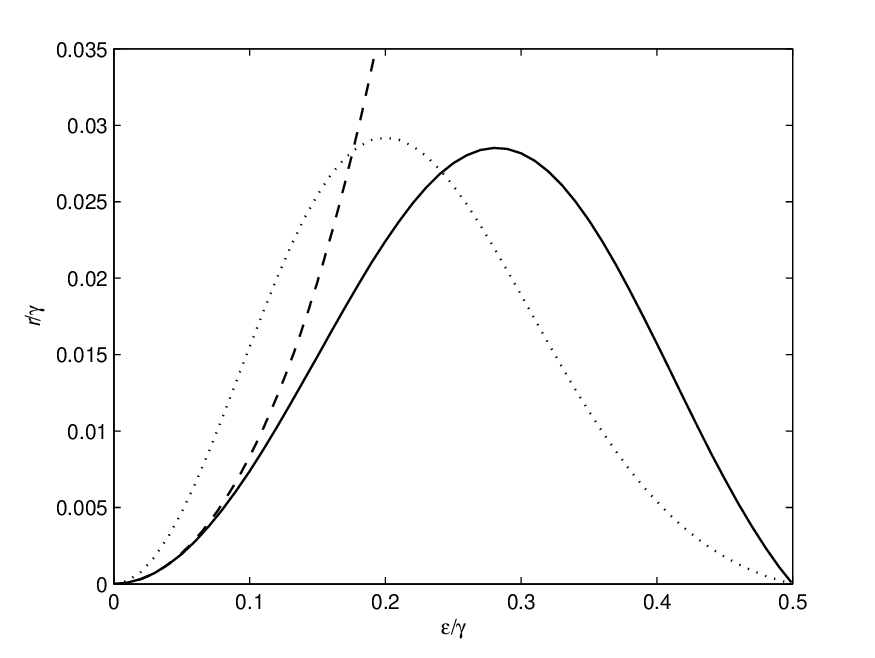}
\caption{Production rate $r$ as a function of $\epsilon$. The dotted
line corresponds to $\eta_t=1$ and $T\gamma=10$, while the solid
line is for $\eta_t=0.4$ and $T\gamma=10$. The dashed line is for
$\eta_t=0.4$ and $T\gamma=0$.} \label{epsr}
\end{center}
\end{figure}

The production rates corresponding to the solid line and the dotted
line in figure \ref{femepsfidid} are shown in figure \ref{epsr}, and
the dashed line in figure \ref{epsr} is the rate for $T=0$ and
$\eta_t=0.4$. The production rate vanishes for zero intensity
because there are no photons present in the trigger beam in that
limit. Also in the large intensity limit the rate approaches zero
for $T\neq0$. The reason is that the large photon flux makes it
extremely unlikely to observe no clicks during some non-zero time
interval. It is desirable to be on the left side of the maximum of
the relevant curve in figure \ref{epsr} because the fidelity is a
decreasing function of $\epsilon$. Hence for $T\gamma=10$ and
$\eta_t=1$, $\epsilon/\gamma$ should be below 0.20, while for
$\eta_t=0.4$, $\epsilon/\gamma$ should be below 0.28. The
$\eta_t=0.4$ curve is shifted to the right compared to the
$\eta_t=1$ curve because a lower detector efficiency corresponds to
a lower photon flux in the beam. The maximal rate
($r/\gamma\approx0.029$ for both curves) corresponds to a mean
temporal distance between accepted clicks of $34\textrm{
}\gamma^{-1}$. For $\eta_t=0.4$ the high fidelity region is
$\epsilon/\gamma\lesssim0.15$. For this region it is apparent from
figure \ref{epsr} that the decrease in production rate for fixed
$\epsilon/\gamma$ is quite small when $T\gamma$ is increased from 0
to 10, while figure \ref{femepsfidid} shows that the fidelity
increase $\Delta F_1$ is significant.

\section{Conclusion}

In conclusion we have considered conditional single photon state
generation from a continuous-wave non-degenerate OPO. We have
presented explicit analytical expressions for the output state
Wigner function, the one-photon state fidelity, the production rate,
and the optimal zero intensity limit temporal mode function for the
signal state. We have optimized the mode function numerically to
achieve maximal fidelity, and from a variational calculation we
determined an integral equation for the optimal mode function. By
conditioning on clicks surrounded by an interval with no clicks, the
fidelity was increased for fixed $\epsilon/\gamma$, while the
production rate was decreased slightly. However, accepting a smaller
fidelity increase by increasing $\epsilon/\gamma$ led to a larger
production rate if $T\gamma$ and $\epsilon/\gamma$ were not too
large.

Using our approach, it is possible analytically to calculate the
output state Wigner function conditioned on arbitrary sequences of
click detections, vacuum (i.e. no click) detections, and no
detections (i.e. detector turned off) in temporal trigger modes of
short duration, and the corresponding production rate can also be
determined analytically since both calculations involve integration
of products of polynomials and Gaussian functions only. From the
conditioned output state Wigner function it is straightforward to
compute the fidelity for arbitrary states in arbitrary output modes.
We are currently working on a generalization of our calculations for
the production of single photon states to generation of $n$-photon
and Schr\"odinger kitten states.

The authors acknowledge discussions with Jonas S. Neergaard-Nielsen
and Eugene S. Polzik, and financial support from the European Union
through the Integrated Project "SCALA".

\appendix

\section{Wigner function and fidelity for one-photon states
produced from a degenerate OPO}

The setup analyzed in \cite{klaus} is obtained from figure
\ref{setup} by replacing the non-degenerate OPO with a degenerate
OPO and the polarizing beam splitter with a normal beam splitter
with low transmission. The two-time correlation functions for this
setup are \cite{klaus}
\begin{eqnarray}\label{cor1}
\langle\hat{a}(t)\hat{a}(t')\rangle&=&\frac{\lambda^2-\mu^2}{4}
\left(\frac{\textrm{e}^{-\mu|t-t'|}}{2\mu}+
\frac{\textrm{e}^{-\lambda|t-t'|}}{2\lambda}\right),\\
\label{cor2}
\langle\hat{a}^\dag(t)\hat{a}(t')\rangle&=&\frac{\lambda^2-\mu^2}{4}
\left(\frac{\textrm{e}^{-\mu|t-t'|}}{2\mu}-
\frac{\textrm{e}^{-\lambda|t-t'|}}{2\lambda}\right),
\end{eqnarray}
and \eqref{modeope} and \eqref{modeope2} are modified to
\begin{eqnarray}\label{mop1}
\hat{a}_1&=&\int
f_1(t')\Big(\sqrt{\eta_t(1-R)}\hat{a}(t')-\sqrt{\eta_tR}\hat{a}_{vac}(t')\nonumber\\
&&\qquad\qquad\qquad+\sqrt{1-\eta_t}\hat{b}_{vac}(t')\Big)\mathrm{d}t',\\
\label{mop2}\hat{a}_2&=&\int
f_2(t')\Big(\sqrt{\eta_sR}\hat{a}(t')+\sqrt{\eta_s(1-R)}\hat{a}_{vac}(t')\nonumber\\
&&\qquad\qquad\qquad+\sqrt{1-\eta_s}\hat{c}_{vac}(t')\Big)\mathrm{d}t',
\end{eqnarray}
where $R$ is the beam splitter reflectivity, and the operators
labeled by $vac$ are field operators acting on vacuum. As in Sec. IV
we can restrict our analysis to real signal mode functions. In terms
of the covariance matrix elements computed from \eqref{cor1},
\eqref{cor2}, \eqref{mop1}, and \eqref{mop2} the Wigner function for
the produced state conditioned on a single trigger detector click
event is
\begin{equation}
W_{click}(x_2,p_2)=\frac{1}{C_1}(C_2+C_3x_2^2+C_4p_2^2)
\mathrm{e}^{-C_5x_2^2-C_6p_2^2},
\end{equation}
where
\begin{eqnarray*}
C_1&=&\pi(V_{33}V_{44})^{5/2}(V_{11}+V_{22}-2),\\
C_2&=&V_{33}V_{44}(V_{33}V_{44}(V_{11}+V_{22}-2)-V_{33}V_{24}^2-V_{44}V_{13}^2),\\
C_3&=&2V_{13}^2V_{44}^2,\\
C_4&=&2V_{24}^2V_{33}^2,\\
C_5&=&(V_{33})^{-1},\\
C_6&=&(V_{44})^{-1},
\end{eqnarray*}
and the one-photon state fidelity is
\begin{multline}
F_1(f_2(t))=\frac{2(V_{33}V_{44}-1)}
{(1+V_{33})^{3/2}(1+V_{44})^{3/2}}+\\
\frac{2(2(1+V_{44})+1-V_{33}V_{44})V_{13}^2}
{(V_{11}+V_{22}-2)(1+V_{33})^{5/2}(1+V_{44})^{3/2}}+\\
\frac{2(2(1+V_{33})+1-V_{33}V_{44})V_{24}^2}
{(V_{11}+V_{22}-2)(1+V_{33})^{3/2}(1+V_{44})^{5/2}}.
\end{multline}

\end{document}